\documentclass[12pt]{iopart}
\usepackage{graphicx}
\def\gap{\;\rlap{\lower 2.5pt
 \hbox{$\sim$}}\raise 1.5pt\hbox{$>$}\;}
\def\lap{\;\rlap{\lower 2.5pt
   \hbox{$\sim$}}\raise 1.5pt\hbox{$<$}\;}
\def\gsim{\;\rlap{\lower 2.5pt
 \hbox{$\sim$}}\raise 1.5pt\hbox{$>$}\;}
\def\lsim{\;\rlap{\lower 2.5pt
   \hbox{$\sim$}}\raise 1.5pt\hbox{$<$}\;}
\def\msun{{M_\odot}}
\def\rsun{{R_\odot}}

\def\cm{{\rm\,cm}}
\def\Mpc{{\rm\,Mpc}}
\def\kpc{{\rm\,kpc}}
\def\pc{{\rm\,pc}}

\def\GeV{{\rm\,GeV}}
\def\TeV{{\rm\,TeV}}
\def\sec{{\rm\,s}}
\def\sr{{\rm\,sr}}

\def\spose#1{\hbox to 0pt{#1\hss}}
\def\lta{\mathrel{\spose{\lower 3pt\hbox{$\mathchar''218$}}
     \raise 2.0pt\hbox{$\mathchar''13C$}}}
\def\gta{\mathrel{\spose{\lower 3pt\hbox{$\mathchar''218$}}
     \raise 2.0pt\hbox{$\mathchar''13E$}}}
\newcommand{\beq}{\begin{equation}}
\newcommand{\eeq}{\end{equation}}
\newcommand{\be}{\begin{equation}}
\newcommand{\ee}{\end{equation}}

\newcommand{\ls}{\mathrel{\raise1.16pt\hbox{$<$}\kern-7.0pt 
\lower3.06pt\hbox{{$\scriptstyle \sim$}}}}         
\newcommand{\gs}{\mathrel{\raise1.16pt\hbox{$>$}\kern-7.0pt 
\lower3.06pt\hbox{{$\scriptstyle \sim$}}}}         
\def\VEV#1{{\langle #1 \rangle}}
\long\def\comment#1{}

\def\msun{M_{\odot}}
\def\fun#1#2{\lower3.6pt\vbox{\baselineskip0pt\lineskip.9pt
  \ialign{$\mathsurround=0pt#1\hfil##\hfil$\crcr#2\crcr\sim\crcr}}}
\def\lap{\mathrel{\mathpalette\fun <}}
\def\gap{\mathrel{\mathpalette\fun >}}
\newcommand{\ba}{\begin{eqnarray}}
\newcommand{\ea}{\end{eqnarray}}

\begin{document}

\title{$\gamma$-ray flux from
Dark Matter Annihilation in Galactic Caustics}

\author{Lidia Pieri$\sharp$\footnote[3]{e-mail: lidia@physto.se}\ and Enzo Branchini\ddag  
}

\address{$\sharp$\ Department of Physics, Stockholm University, AlbaNova
University Center \\
SE-10691 Stockholm, Sweden
}

\address{\ddag\ Department of Physics, Universit\`a di Roma Tre\\
Via della Vasca Navale 84\\
I-00146 Rome, Italy
}

\begin{abstract}
In the frame of indirect dark matter searches we investigate the flux
of high-energy $\gamma$-ray photons produced by 
annihilation of dark matter in caustics within our Galaxy
under the hypothesis that the bulk of dark matter is composed of the
lightest supersymmetric particles.
Unfortunately, the detection of the caustics annihilation signal
with currently available instruments is rather challenging. 
Indeed, with realistic assumptions concerning particle physics and cosmology,
the $\gamma $-ray signal from caustics is 
below the detection threshold of both $\check {\rm C}$erenkov telescopes and 
satellite-borne experiments.
Nevertheless,
we find that this signal is more prominent than that 
expected if annihilation only occurs in the smoothed Galactic halo,
with the possible exception of a 
$\sim 15^{\circ}$ circle around the Galactic center 
if the mass density profile of our Galaxy exhibits a sharp cusp there.
We show that the angular distribution of this $\gamma$-ray flux
changes significantly if DM annihilation preferentially occurs
within virialized sub-halos populating our Galaxy
rather than in caustics.
\end{abstract}



\section{Introduction}

Unveiling the nature and distribution of the dark matter (DM) 
is a fundamental goal of both theoretical and experimental physics today.   
The latest observations \cite{concordance} allow  
$\sim 23\%$ of the matter in the universe to be of unknown,
non-baryonic nature.
One of the strongest candidates for the DM is a weakly interacting massive
particle 
that was non-relativistic at the epoch of decoupling from the rest of the
universe, as required by the currently
popular cold dark matter (CDM) cosmological model. 
\noindent
A good DM candidate is represented by
the lightest supersymmetric particle (LSP)
that in most particle physics
scenarios is the neutralino $\chi$. This particle is a spin-$\frac{1}{2}$ 
Majorana fermion constituted of a linear combination of the neutral gauge 
bosons and neutral Higgs doublet spartners, 
$\chi = a \tilde B + b \tilde{W_3} + c \tilde{H_1^0} + d \tilde{H_2^0}$.
In the popular SUGRA or SUGRA-like models, 
where gaugino-universality is required, 
its mass is constrained by LEP limits to be greater than $\sim 50 \GeV$
\cite{susylep2} while theoretical considerations place an upper
limit of a few hundred GeV - few TeV \cite{Ellis:00,Jungman:96} to the 
neutralino mass.
If $R$-parity is conserved, neutralinos can change their cosmological 
abundance only through annihilation. Hence, high density regions 
constitute preferential locations for indirect DM searches. \\

Most of the recent works have assumed that the bulk of DM is
constituted of neutralinos
and looked for a possible detection of $\gamma$-ray annihilation flux with
appropriate detectors. 
Phenomenological approaches to DM searches 
have focused mainly on the Galactic Center (GC) that has been
observed by the 
EGRET satellite in the energy range 100 MeV - 10 GeV \cite{egret_gc} and,
more recently, by the ground-based $\check {\rm C}$erenkov telescopes
WHIPPLE \cite{WHIPPLE}, CANGAROO-II \cite{CANGAROOGC} and HESS \cite{hessGC}
with an energy threshold of hundreds of GeV.
These data have shown an enhanced  $\gamma$-ray flux 
along the the GC direction
that has been variously interpreted and, although puzzling, does not 
yet constitute a firm evidence of indirect DM detection.  
New data from VERITAS \cite{VERITAS} and MAGIC \cite{MAGIC} 
$\check {\rm C}$erenkov telescopes, as well as from the soon-to-come satellite
experiment GLAST \cite{GLAST}, will certainly help clarifying this issue. \\

If dark relics are stable and their annihilation cross-section is
non-negligible, then regions of enhanced density constitute preferential 
locations for the annihilation of DM particles.
Our aim is to quantify the DM annihilation rate
and to predict the associated $\gamma$-ray flux occurring 
within those singularities that arise in irrotational flows in a cold, 
collisionless medium during the process of structure formation,  
usually referred to as ``caustics'' because of
their analogy with caustics in geometrical optics \cite{ZMS:83}.
Although the presence of DM caustics is a 
general prediction of CDM cosmologies, model details 
depend on the characteristics of the DM candidate and 
on our ability in following the build up of caustics in the very nonlinear
dynamical regime \cite{CR,One,VP,HWS}.
In this work we will assume that the DM is composed of neutralinos and
that  the  shape and position of Galactic caustics can be modeled 
as in \cite{VP}. \\

The rationale behind this work is to predict the 
annihilation signal from DM caustics within our Galaxy,
characterize its spatial signature and 
evaluate whether currently available or next-generation $\gamma$-ray detectors 
will be able to detect this signal and discriminate it from the 
$\gamma$-ray Galactic background. 
Our results will be compared with those obtained from a previous
similar analysis performed by \cite{BEG,GT} that,
however, considered the case of a single, nearby caustic of non-negligible 
transverse size. \\

The centers of virialized DM halos also constitute possible sites
for annihilation of DM particles if a singularity in the mass density field 
called ``cusp' develops there as a result of the gravitational instability 
process.
Theoretical considerations and numerical experiments have neither been able
to set strong constraints on the shape of the density profile near the
halo center \cite{Navarro:97,Moore:99,ullio:01,Navarro:03,Moore:04,Power:03,stoehr:03} 
nor to predict the abundance of
a population of dark galactic sub-halos 
and their survival time within the Galactic halo
\cite{Hayashi:02,helmi:03,gao:04,DM:05,ZT:05}.
On the other hand,current observations such as
the rotation curves in the central galactic regions
\cite{deBlok:03,swaters:03,donato:04}, 
Planetary Nebulae kinematics \cite{merrifield:04},
microlensing events toward the Galactic bulge
\cite{afonso:03}
and spectroscopic gravitational lensing \cite{metcalf:04}
do not allow yet to discriminate among the various scenarios,
although they seem to favor a very shallow, or cored 
density profile.
Nevertheless, many authors have considered the
possibility of detecting the products of DM annihilation occurring 
within Galactic, extragalactic and sub-galactic DM halos, 
taking into account both observational
and theoretical constraints 
\cite{FPS:04,gamma_papers,stoehr:03,lightindirect, 
PB:04,Roldan:00,extrapapers}. \\

As neither theoretical arguments nor numerical experiments
have been able yet to construct self consistent dynamical models
that account for both DM caustics and virialized substructures
populating a host DM matter halo  it is difficult
to quantitative compare the annihilation flux produced
within either structures. 
Nevertheless, as we discuss in this work,
qualitative differences between the angular distribution
of the signals exist that may help in 
discriminating among these two scenarios. \\
 
The layout of the paper is as follows.
In Section 2 we describe the expected photon flux from neutralino annihilation
and introduce the detectability parameters for 
${\rm \check{C}}$erenkov telescopes and satellite-borne experiments.
In Section 3 we study
the effect of a population of DM 
caustics  on the expected $\gamma$-ray flux.
Section 4 is devoted to investigating the possibility of detecting
high energy photons produced in DM caustics.
In Section 5 we summarize and discuss our main conclusions.

\section{Detecting $\gamma$-rays from neutralino annihilation} 

Following the prescriptions given in \cite{FPS:04}, we factorize
the diffuse photon flux from neutralino annihilation as
\begin{equation}
\frac{d \Phi_\gamma}{dE_\gamma}(E_\gamma, \psi, \Delta \Omega) =
\frac{d \Phi^{\rm SUSY}} {dE_\gamma}(E_\gamma) \times 
\Phi^{\rm cosmo}(\psi, \Delta \Omega),
\label{flussodef}
\end{equation}
where $\psi$ is the angle of view from the GC and 
$\Delta \Omega$ is the solid angle of observation for the detector.
We will not consider here the $\gamma$-line emission. 
Its branching ratio is in fact as small as $10^{-3}$ and it produces
a very small photon flux which we will neglect.
In Eq. \ref{flussodef} the particle physics is embedded in the term:
\begin{equation}
\frac{d \Phi^{\rm SUSY}}{dE_\gamma}(E_\gamma) =  
  \frac{1}{4 \pi} \frac{\VEV{\sigma_{\rm ann} v}}{2 m^2_\chi} \cdot 
\sum_{f} \frac{d n^f_\gamma}{d E_\gamma} b_f  
\label{flussosusy}
\end{equation}
where $\VEV{\sigma_{\rm ann}v}$ is the thermally-averaged 
neutralino self-annihilation cross-section times the relative 
velocity of the two annihilating particles, 
$d n^f_\gamma / dE_\gamma$ is the differential photon
spectrum for a given annihilation final state $f$, whose analytical
expression is given in \cite{FPS:04},
$b_f$ is the branching ratio of annihilation into $f$
and $m_\chi$ is the neutralino mass. \\
Astrophysics and cosmology affect the last factor of Eq. \ref{flussodef}, 
that represents the line-of-sight integral:
\begin{equation}
\Phi^{\rm cosmo}(\psi, \Delta \Omega) = \int_{\Delta \Omega}
d \Omega' \int_{\rm l.o.s} \rho_{\chi}^2 (r(\lambda,\psi')) d\lambda(r,\psi'),
\label{flussocosmoMW}
\end{equation}
where 
$\rho_\chi(r)$ is the neutralino density profile, $r$ is 
the distance from the GC and is equal to
$r = \sqrt{\lambda^2 + \rsun^2 -2 \lambda \rsun \cos \psi}$, 
$\lambda$ is the distance from the observer and $\rsun$ is the 
distance of the Sun from the GC.

For distant, extragalactic objects, Eq.~\ref{flussocosmoMW} reads
\begin{equation}
\Phi^{\rm cosmo}(\psi, \Delta \Omega) =  
\frac{1}{d^2} 
\int_{0}^{\min[R_G,r_{\rm max}(\Delta \Omega)]}  
4 \pi r^2 \rho_{\chi}^2(r) dr, 
\label{flussocosmoextra}
\end{equation}
where 
$d$ is the distance of the object from the observer, 
$R_G$ is its virial radius and $r_{\rm max}(\Delta \Omega)$ 
is the radius seen within the angular acceptance of the detector. \\

As mentioned before, the mass distribution near the center
of the DM halo is poorly constrained.
High resolution numerical experiments have shown that a 
cuspy density profile develops near the halo centers
with an inner slope that is bracketed between the
NFW profile (hereafter NFW97) \cite{Navarro:97}
\begin{equation}
\rho_\chi^{\rm NFW97} = \frac{\rho^{\rm NFW97}_s}{\left (r/r^{\rm NFW97}_s \right )
\left (1 + r/r^{\rm NFW97}_s \right )^{2}}
\label{eqnfw}
\end{equation}
and the steeper Moore et al. profile (M99) \cite{Moore:99}:
\begin{equation}
\rho_\chi^{\rm M99} = \frac{\rho_s^{\rm M99}}{\left (r/r_s^{\rm M99} 
\right )^{1.5} 
\left [1 + \left (r/r_s^{\rm M99}\right )^{1.5} \right ]}.
\label{eqmoore}
\end{equation}
Scale quantities are fixed by the observations (virial mass or 
peak rotation velocity of the halo) and by the estimation
of the concentration parameter $c = r_{\rm vir}/r_s$, where we  
define the virial radius $r_{\rm vir}$ as the radius
within which the halo average density is 200 times the critical
density of the universe. The parameters
$c_{\rm NFW97}$ and $c_{\rm M99}=0.64 \
c_{\rm NFW97}$ have been computed according to \cite{ENS} assuming a 
CDM universe.

As discussed in Section 1, the existence of a central cusp has not been 
confirmed by recent astronomical observations that, instead, seem to 
favor the case of a central core of constant density.
Cored density profile would significantly depress the 
annihilation rate in the central region \cite{evans:04}.
However, since using a cored rather than a cuspy density profile 
would not significantly affect our flux predictions
for  $\psi>15^{\circ}$, where the caustics signal is expected to be 
more prominent, in this work we only consider the  M99 and NFW97 
model density profiles. 

\begin{figure} 
\includegraphics[height=8cm,width=8cm]{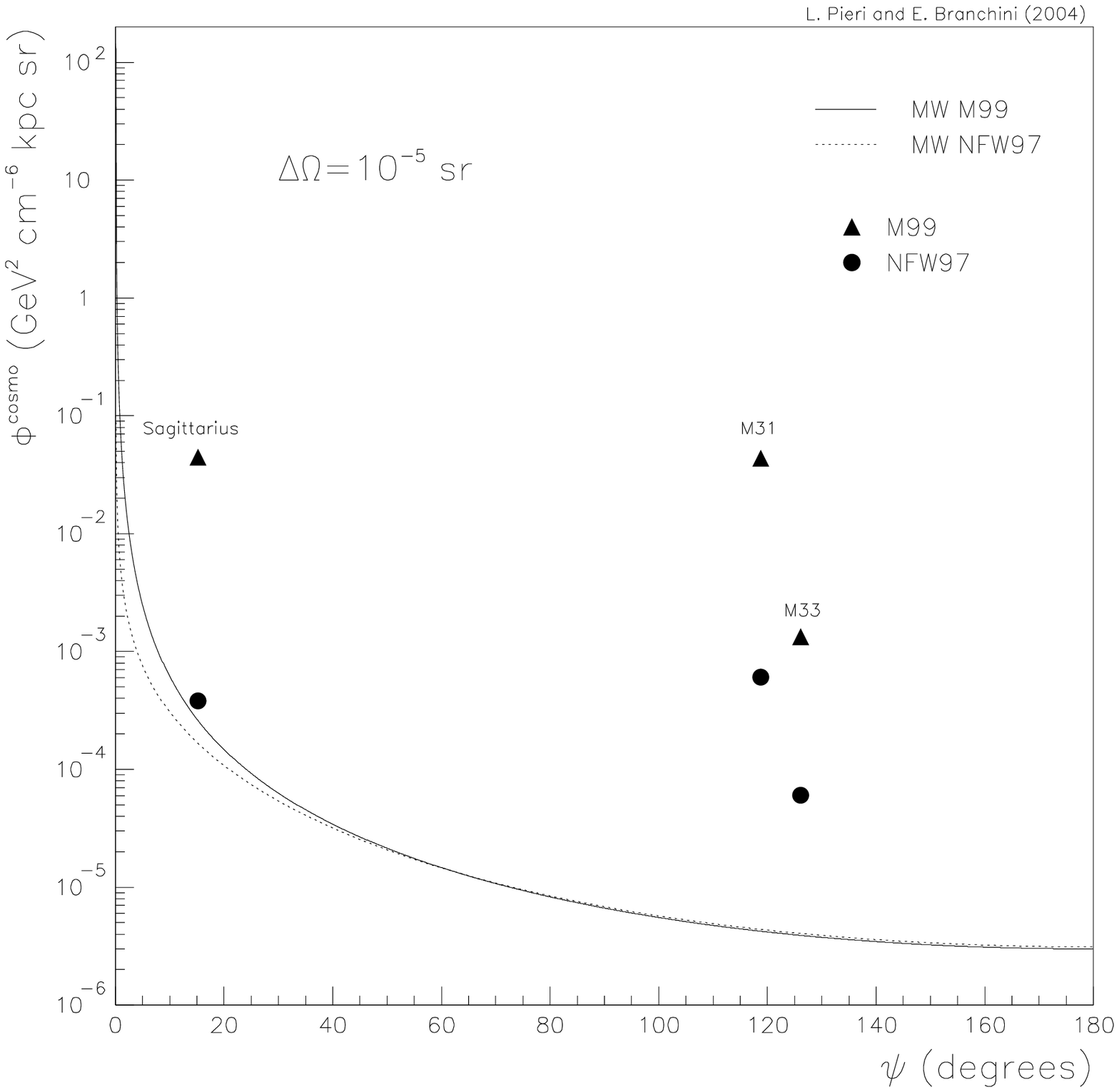}
\includegraphics[height=8cm,width=8cm]{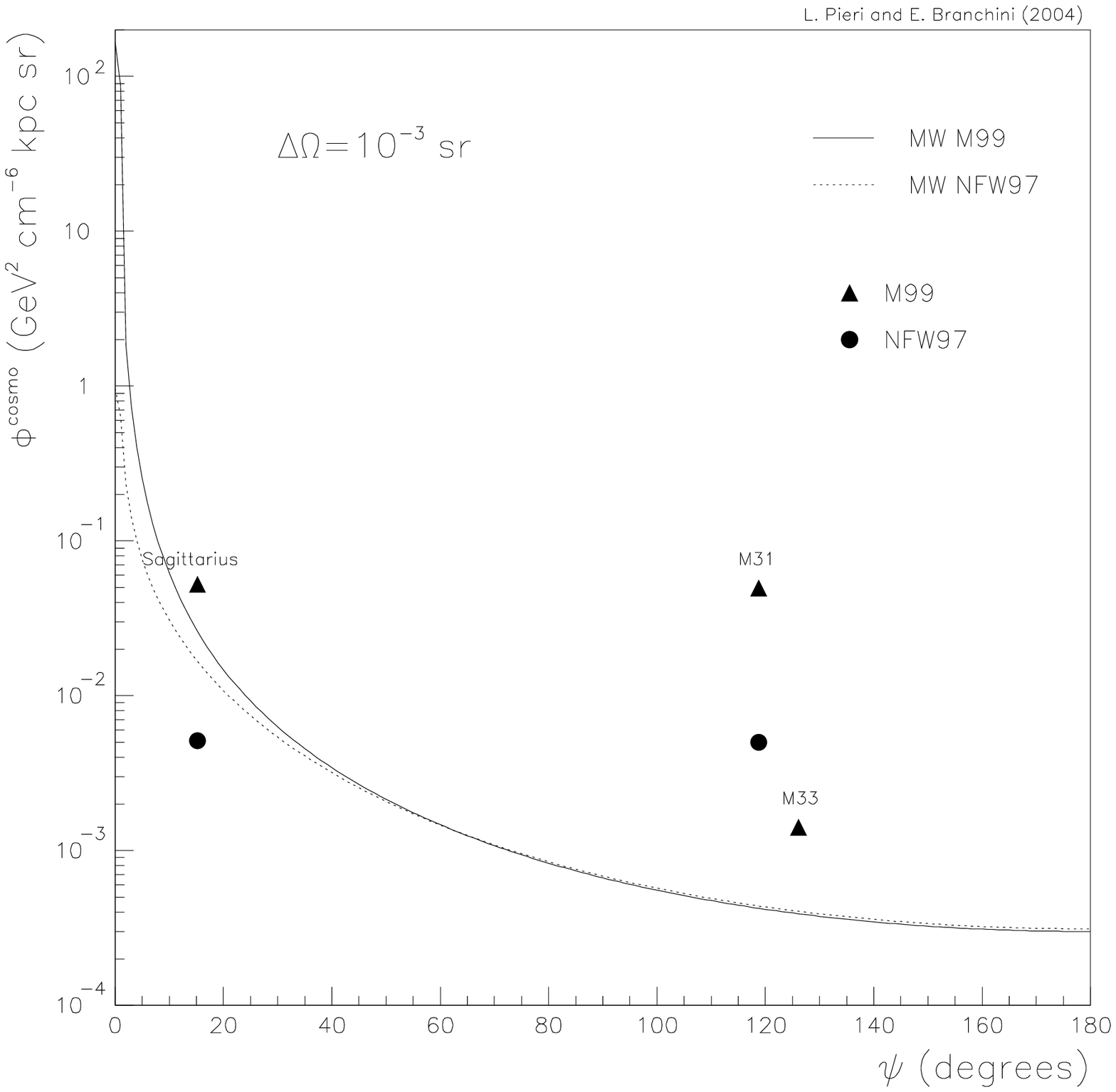}
\caption{``Cosmological factor'' $\Phi^{\rm cosmo}(\psi)$ for the expected 
$\gamma$-ray flux
from the most prominent external galaxies and for the Milky Way 
foreground within a solid angle  
$\Delta \Omega=10^{-5} sr$ (left panel) or $\Delta \Omega=10^{-3} sr$ 
(right panel), for a M99 (solid line and triangles) and
a NFW97 (dotted line and circles) profile. A small central core region
of radius $10^{-8} \kpc$ has been assumed. 
}
\label{fig1}
\end{figure}

Fig.\ref{fig1} shows the factor $\Phi^{\rm cosmo}(\psi)$
of Eq. \ref{flussocosmoMW} for our Galaxy and  
the three more prominent galaxies
within a distance of $1.5 \Mpc$,
computed for a solid angle of $10^{-5} \sr$ (left panel) 
and $10^{-3} \sr$ (right panel) assuming both a NFW (filled dots) and a Moore 
profile (filled triangles). We would like to remark here that the use of
a cored profile would substancially reduce the expected fluxes. \\
Divergences in the l.o.s. integral 
are avoided by forcing a constant density core 
of radius $10^{-8} \kpc$ corresponding to the distance at which
the self-annihilation rate equals the dynamical time.
Several astrophysical effects can increase the size of
this core radius, leading to a significant decrease in the
annihilation rate and expected flux
\cite{PB:04,FPS:04}. 
As already pointed out by \cite{PB:04}, these external 
galaxies shine above the Galactic foreground 
as clearly seen in both panels of Fig.\ref{fig1}.

The ``supersymmetric factor'' $\frac{d \Phi^{\rm SUSY}} {dE_\gamma}$
in Eq. \ref{flussosusy} has been investigated   extensively by \cite{FPS:04}. 
In our calculations we assume 
$\Phi^{\rm SUSY} = 10^{-32}  
\cm^3 \sec^{-1} \GeV^{-2} \sr^{-1}$ which represents 
a reasonable though somewhat optimistic choice among the 
values allowed by  supersymmetric models. \\

\subsection{Experimental detectability}\label{subsec:det}
One of the aims of this work is to study the experimental detectability of
DM $\gamma$-ray sources with both ${\rm \check{C}}$erenkov telescopes
and satellite-borne experiments.
Following \cite{FPS:04,PB:04}, we explore the 
detectability of the diffuse photon flux from DM annihilation by
comparing the number $n_\gamma$ of expected $\gamma$ events from 
neutralino annihilation with the
fluctuation of background events $n_{\rm bkg}$:
\begin{equation}
\sigma \equiv \frac{n_{\gamma}}{\sqrt{n_{\rm bkg}}} =
\frac{\sqrt{T_\delta} \epsilon_{\Delta \Omega}}{\sqrt{ \Delta \Omega}}
\frac{\int A^{\rm eff}_\gamma [d\phi^{\rm DM}_\gamma/dE
d\Omega] dE d\Omega}{\sqrt{ \int \sum_{\rm bkg} A^{\rm eff}_{\rm
bkg} [d\phi_{bkg}/dEd\Omega] dE d\Omega}}. 
\label{sensitivity}
\end{equation}
To estimate the experimental sensitivity, $\sigma$, we choose the subsequent,
 rather typical set of  parameters. 
For the effective observation time, $T_\delta$,
we consider a 20-day pointing for 
${\rm \check{C}}$erenkov telescope 
and a 30-day observation for the satellite-borne experiment 
(for a ${\rm \check{C}}$erenkov telescope, only periods corresponding to
a source zenith angle $\theta \leq 60^{\circ}$ are considered). 
The fraction of signal events $\epsilon_{\Delta \Omega}$ is set equal to $0.7$ 
within the optimal solid angle $\Delta \Omega = 10^{-5} \sr$, 
corresponding to an angular resolution of the instrument of $0.1^{\circ}$. 
The effective detection areas for electromagnetic and hadronic induced showers 
are considered to be independent from the value of the particle
energy and incident angle and are set equal to
 $A^{\rm eff}_{\rm \check{C}erenkov} = 4 \times 10^8 \cm^2$ and
$A^{\rm eff}_{\rm satellite} = 10^4 \cm^2$, respectively.
The identification efficiency $\epsilon$ is of crucial importance 
to reduce the physical background level.  A ${\rm
\check{C}}$erenkov telescope has a typical identification efficiency
$\epsilon_{\rm e.m.}\sim 99 \%$
for electromagnetic induced (primary $\gamma$ or electrons) showers
 and $\epsilon_{\rm had}\sim 99 \%$ for hadronic showers.
A satellite experiment has an identification efficiency for charged
particles of $\epsilon_{\rm charged}\sim 99.997 \%$
and of $\epsilon_{\rm neutral}\sim 90 \%$ for photons, due
to the backslash of high energy photons \cite{moiseev}.
Our expected signal is compared to the presence
of proton \cite{hbck},
electron \cite{elbck} and photon \cite{Bergstrom:98,gbck} backgrounds.

\section{The caustics scenario}

The caustics phenomenon has been extensively studied by 
\cite{causticpapers}. Here
we briefly outline the main characteristics of the DM caustics that
are relevant for our work.

Collisionless dark matter particles move in the universe on a thin sheet 
in the velocity-position phase space, whose thickness
is determined by their small primordial velocity dispersion.
For a neutralino such a velocity has been calculated, 
after the equivalence time and before the onset of galaxy formation, 
to be as small as $\delta v_\chi (t) \sim 10^{-11} c \ {\rm(GeV/m_\chi)^{1/2} 
(t_0/t)^{2/3}}$, where ${\rm t_0}$ is the age of the universe. Such a sheet
winds up where an overdensity grows non-linearly (i.e. when a galaxy forms)
and the particles on the sheet fall onto the corresponding overdensity. 
The distance from the center of the overdensity at which the folding occurs 
has been estimated to be of the order of $1 \Mpc$ for a 
galaxy like the Milky Way.

The following period of continuous cross-infalls lasts 
until inhomogeneities diffuse the process 
and prevent the formation of 
structures. In this period, a set of caustic surfaces which wrap around
the galaxy is formed where the phase space folds.  
These surfaces define areas of enhanced density located
at the boundary between regions characterized by different numbers 
of flows of particles.
Namely, in the internal region of the caustic surface 
there is an odd number $n$ of streams of particles compared to
the $n-2$ flows in the external region.
The non-zero velocity dispersion prevents the mass density to
diverge at the caustics surface. 
More precisely, the density at the caustics is smeared off on distances
$\delta x_\chi \sim 3 \cdot 10^{15} {\rm \ cm \ (GeV/m_\chi)^{1/2}}$, 
corresponding to about $10^{-4}$ pc for a TeV-mass neutralino.
The final outcome of this process of caustics formation is the
onset of a series of caustics with different characteristics. 
A number of {\it outer} caustics form, that appear as 
topological spheres surrounding the galaxy, as well as
{\it inner} caustics, which look like tubes 
lying in the Galactic plane, composed of the 
particles which carry the most angular momentum when they are at their
distance of closest approach to the Galactic Center. 
At the intersection between the Galactic plane and a 
perpendicular plane passing through the GC, these caustic 
tubes have a triangular shaped 
cross-section  with three density cusps at the vertices of the 
triangle (the so-called tricusp), one on the Galactic plane 
and the other two located at the extremes of a 
segment perpendicular to it.

\subsection{Density distribution near caustics}

The particle density at the caustics can be calculated analytically \cite{CR}.
In proximity of the cusps of the triangle the  density profile $d^{cau}$
scales like $r^{-1}$ while near both the surface of the outer caustics and
the side of the triangle it scales like $r^{-1/2}$,
with $r$ being the distance from the caustics.
A set of simplifying assumptions is usually made to model the
density distribution in these regions. \\
The claimed observational evidence of a Galactic tube characterized by
a triangle with side of $100 \pc$~\cite{CR} is at variance with 
the theoretical prediction of $1 \kpc$ for the side of the 
triangle of the nearest inner caustics, suggesting that neither 
observations nor theory are currently able to set precise constraints on the 
size of caustic tubes. Therefore, we make the 
assumption that for all practical purposes the 
caustic triangles are of negligible sizes, i.e
that the inner caustics reduce to caustic rings, or lines,  
on the Galactic plane. 
Such an approximation is valid~\cite{CR} as long as the distance
of the observer from the caustic ring is of the same order as
the ring radius. This means that, assuming a galactocentric distance
of $8.5 \kpc$, caustic rings with radii smaller than about 4-5 kpc can 
be considered as lines. 
Yet, the approximation still holds for caustic rings which are closer to
the observer, provided that their transverse dimensions  
are smaller than the distances from the observer  
which, in this case, are smaller than the ring radius.
In the model proposed by~\cite{VP} the nearest caustics is characterized
by a radius of $7.8 \kpc$ and transverse dimensions of $\sim 200 \pc$.
When compared to the galactocentric distance,
we have that $200 < 700 < 7.800  \pc$, i.e. 
both the above conditions are satisfied and the approximation
is valid. 
The previous relation is also valid for the remaining caustics that 
have similar transverse dimension but have larger radii.
To compute the  profile of the $\gamma$-ray emission
from neutralino annihilation in the caustics we 
furthermore assume that the turnaround sphere is initially rigidly 
rotating and that the velocity of each particle in the flow is constant.
Under these hypotheses we obtain a simple expression for the density 
distribution near the caustics.\\
For the the n-th outer caustics we have \cite{One,causticpapers}:
\begin{equation}
d_n(r)=2\frac{dM}{d\Omega dt}\Bigg|_{n}\,\,\frac{1}{r^2
v_{\rm{rot}}\sqrt{2\ln{\left(\frac{R_n}{r}\right)}}},
\label{eq:outercaustic}
\end{equation}
where $r$ is the distance from the GC, $v_{\rm{rot}}$ is
the rotational velocity of the galaxy, $R_n$ is the radius of
the outer caustics 
and $dM$/$d\Omega dt$ is the rate at which mass falls 
in per unit time and unit solid angle, that can be estimated
using the self-similar infall model (see \cite{One} for detailed 
references). \\
The density profile near the m-th caustic ring is given 
by \cite{One,causticpapers}:
\begin{equation}
d(a_m;\rho,z) \simeq {dM \over d\Omega dt}\Bigg|_{m}~ {2 \over v}~
{1 \over \sqrt{(r^2-a_m^2)^2 + 4 a_m^2 z^2}}
\label{eq:innercaustic}
\end{equation}
where $(\rho,z,\theta)$ are cylindrical coordinates with respect to the 
GC, $a_m$ is the caustics
ring radius \cite{VP}, 
$r= \sqrt{\rho^2 + z^2}$
and $v$ is the velocity amplitude of the DM particles
at the caustics position. \\

\subsection{Expected $\gamma$-ray flux from caustics}

\begin{figure}
\begin{center}
\includegraphics[height=7.8cm,width=7.7cm]{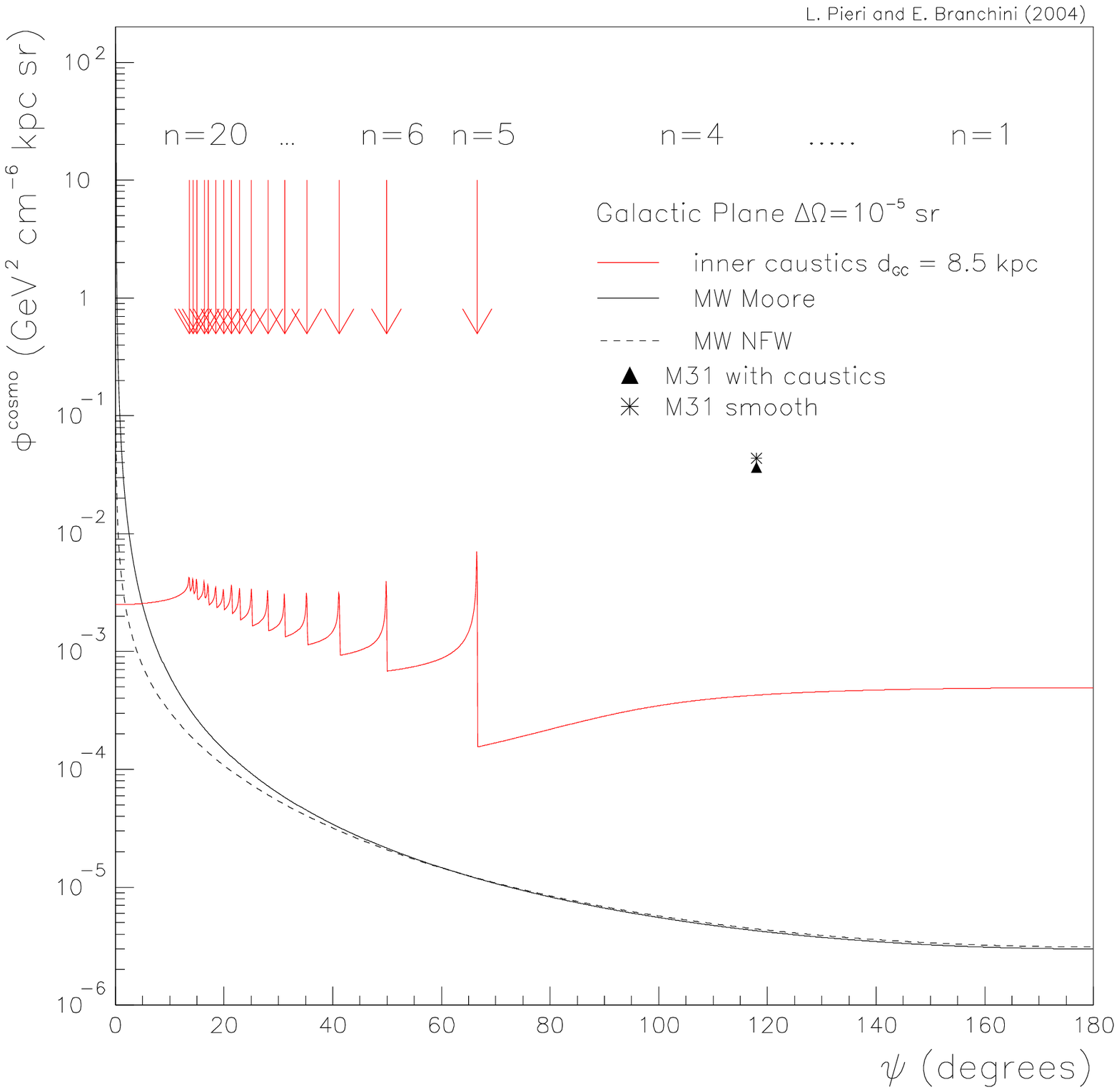}
\includegraphics[height=7.8cm,width=7.7cm]{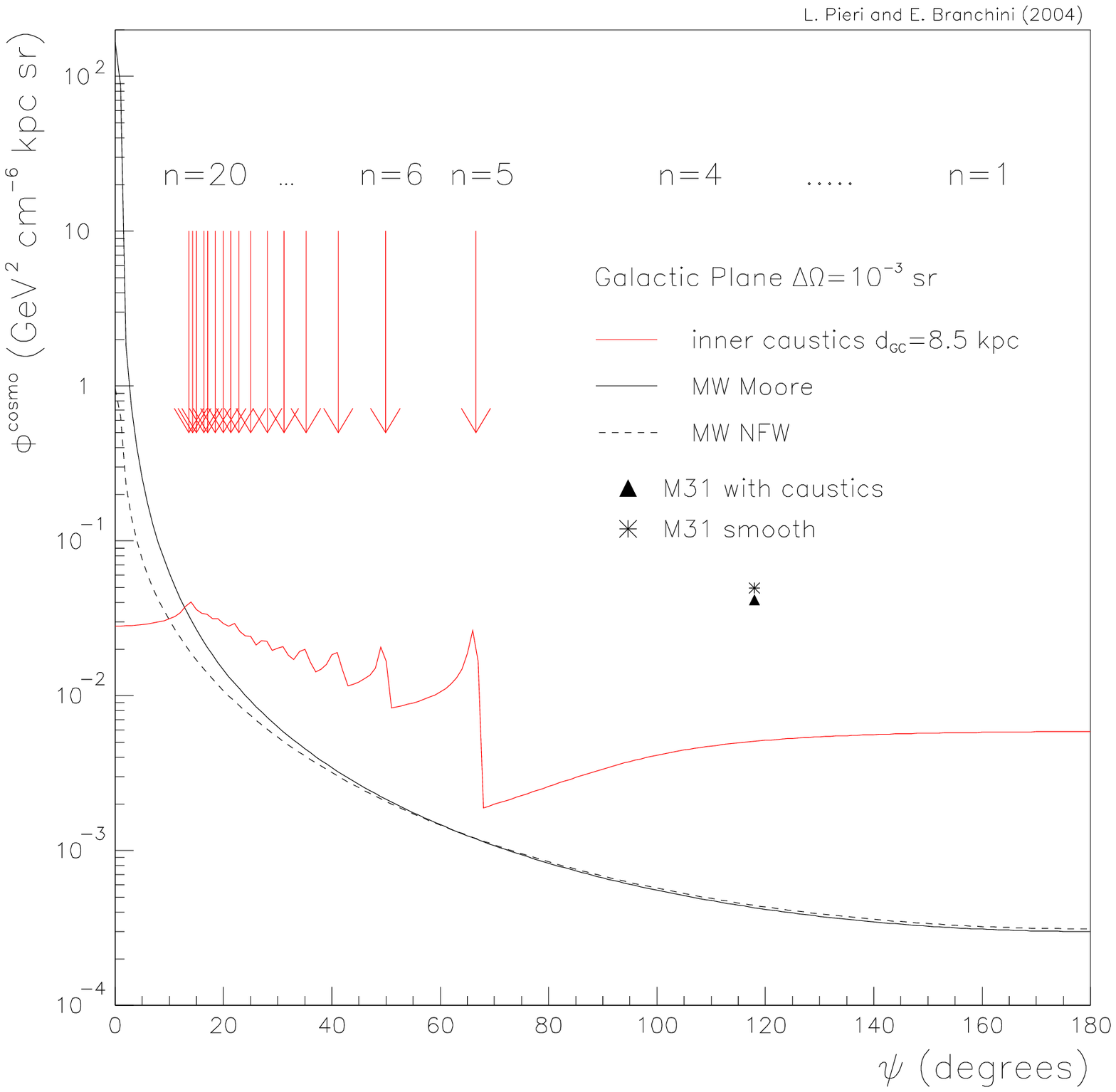}
\end{center}
\caption{``Cosmological factor'' for
the expected $\gamma$-ray emission from neutralino
annihilation in the Galactic plane due to the presence of caustics rings for
an angular resolution of 0.1 degrees corresponding to $10^{-5}$ sr
(left panel) and 1 degree corresponding to  $10^{-3}$ sr (right panel). 
The effect of the presence of caustics in the M31 galaxy is shown as well
(symbols). 
\label{fig:caustics}}
\end{figure}

Eqs.~\ref{eq:outercaustic} and ~\ref{eq:innercaustic} can be used to
compute the expected $\gamma$-ray flux from the caustics in
our Galaxy. 
The resulting value of the ``cosmological factor'' $\Phi^{\rm cosmo}(\psi)$
computed at the Galactic plane is shown in Fig.~\ref{fig:caustics} 
for a solid angle of $10^{-5}$ sr (left panel) and  $10^{-3}$ sr 
(right panel).
The expected flux from caustics is obtained by integrating
Eq.~\ref{flussocosmoMW} over a region in which the mass density, given
by either Eqs. ~\ref{eq:outercaustic} or ~\ref{eq:innercaustic}, exceeds 
that of the underlying smooth Galactic halo.
The mass within caustics, hence the amplitude of the annihilation signal,
only depends on the virial mass of the host halo (or its
rotational velocity). Indeed, in the self-similar infall model 
used to derive Eqs. ~\ref{eq:outercaustic} and ~\ref{eq:innercaustic}, 
the value of $v_{rot}$ determines both the rate of infall mass and 
caustics' distances from the center of the halo.

The contribution from the outer caustics is negligible on the scale
of the plots. A most remarkable spatial modulation signature is present   
which is due to the presence of the inner caustics rings ($n \ge 5$)
with an associated signal well above the Galactic foreground level
for both  the 
M99 (straight line) and the NFW97 (dashed line) model density profiles. 

The contribution of the caustics to the annihilation signal is 
negligible for extragalactic objects. In Fig.~\ref{fig:caustics} 
we show  the case of M31 (filled triangle and starred symbol),
for which the effect of concentrating 
part of the halo mass  in caustics actually decreases the  flux.
This derives from having placed a sizable fraction of 
the M31  mass in the outer caustics falling outside
the viewing angle of the instrument.

\begin{figure}
\begin{center}
\includegraphics[height=8cm,width=8cm]{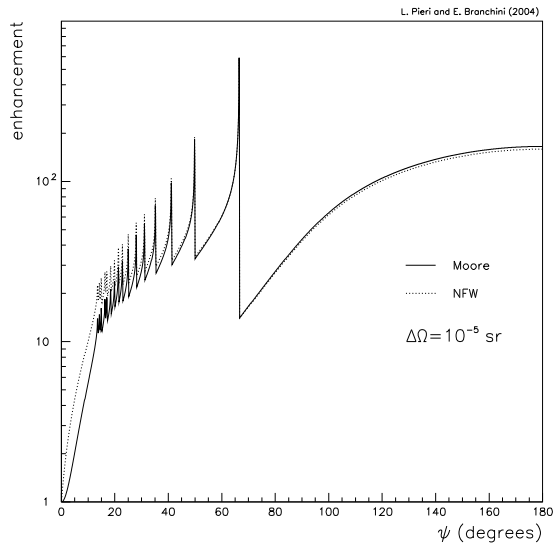}
\end{center}
\caption{Enhancement of the $\gamma$-ray emission from neutralino
annihilation in the Galactic plane due to the presence of caustics, for a
solid angle of $10^{-5}$ sr.
\label{fig:enhcau}}
\end{figure}

Fig.~\ref{fig:enhcau} shows the ratio ($\phi^{MW}+\phi^{cau}$)/$\phi^{MW}$,
which defines the enhancement of the $\gamma$-ray
signal due to the presence of the caustics rings, for a M99 (straight line)
and a NFW97 (dotted line) profile.
As expected, no enhancement is found in the direction of the GC, where
the halo emissivity is very large,
while DM annihilation in the nearest caustics ring 
results in an enhancement factor
of about 600 along the direction $\psi=60^\circ$.
The spatial signature of the $\gamma$-ray flux 
expected in presence of Galactic caustics is
indeed rather prominent.

\subsection{Ring vs. Tricusp}

As we have anticipated, neither theory nor observations are able to set 
strong constraints on the actual size of the caustics that
ranges from $\sim 100$ pc to $\sim 1.0$ kpc.
It is therefore worth investigating how our model predictions change when 
dropping our hypothesis of a ring caustics with negligible transverse 
siize.
This is precisely the case studied by~\cite{BEG} that considered the 
case of the nearest inner caustics (with n=5) located at 7.8 kpc from the 
Galactic Center with transverse dimensions of 1.2 kpc. 
With this particular geometry the Earth, placed at 8.5 kpc from the GC, 
is located well inside the tricusp.

The model~\cite{BEG} also assumes that $m_\chi=100 \GeV$ instead of 
our fiducial value of $1 \TeV$, resulting in a larger smearing distance $\delta x_{\chi}$
and in a smaller cut-off on the caustic density profile 
of $\sim 800 \GeV \cm^{-3}$. Decreasing the density cutoff
reduces the expected annihilation flux.
The reduction factor weakly depends on the geometry of the problem.
In the case of a sigle caustic ring with negligible size,
he value of $\Phi^{MAX}_{\rm cosmo}$ decreases by a factor
of $\sim2$, where $\Phi^{MAX}_{\rm cosmo}$ is the maximum of
$\Phi_{\rm cosmo}(\psi, \Delta\Omega=10^{-5} \ {\rm sr}^{-1})$. 
In the case of a tricusp geometry considered by~\cite{BEG}
the effect is very similar and $\Phi^{MAX}_{\rm cosmo}$
decreases by a factor of $\sim3.3$.

Varying the geometry  while keeping $m_\chi$ constant has a much 
more dramatic impact. We have found that with $m_\chi=100 \GeV$ the value of 
$\Phi^{MAX}_{\rm cosmo}$ for a tricusp is $\sim 100$ smaller than that
expected from a caustic ring with negligible transverse size.
This effect is hardly surprising as in the tricusp scenario, and unlike in the 
ring case, only a small portion of the caustics is seen within the angle of view
$\Delta \Omega$. \\

Overall, the value of  $\Phi^{MAX}_{\rm cosmo}$ for 
the caustics $n=5$ computed as in Section 3.2
is expected to be $\sim 100$ times larger than that of~\cite{BEG}.
Note, however, that discrepancy in the flux predictions of the two 
models are less severe. This derives from the fact that our  value of 
$\Phi^{\rm SUSY} = 10^{-32}  \cm^3 \sec^{-1} \GeV^{-2} \sr^{-1}$
is $\sim 1000$ smaller than that assumed by \cite{BEG}
which, however, requires very optimistic hypotheses about the underlying 
particle physics.

\section{Possibility of $\gamma$-ray flux detection}

To investigate the possibility of detecting the flux of 
$\gamma$-ray photons produced by neutralinos 
annihilating in DM caustics, we compute
the sensitivity of both a ${\rm \check C}$erenkov 
ground-based telescope and a satellite-borne experiment to 
the annihilation fluxes shown 
in Fig.~\ref{fig:caustics}.
The parameters used to compute the experimental sensitivity
(Eq. \ref{sensitivity}) are specified in Section
\ref{subsec:det} and represent realistic observations 
that could be carried out by next-generation detectors.
 
\begin{figure}
\begin{center}
\includegraphics[height=7.8cm,width=7.7cm]{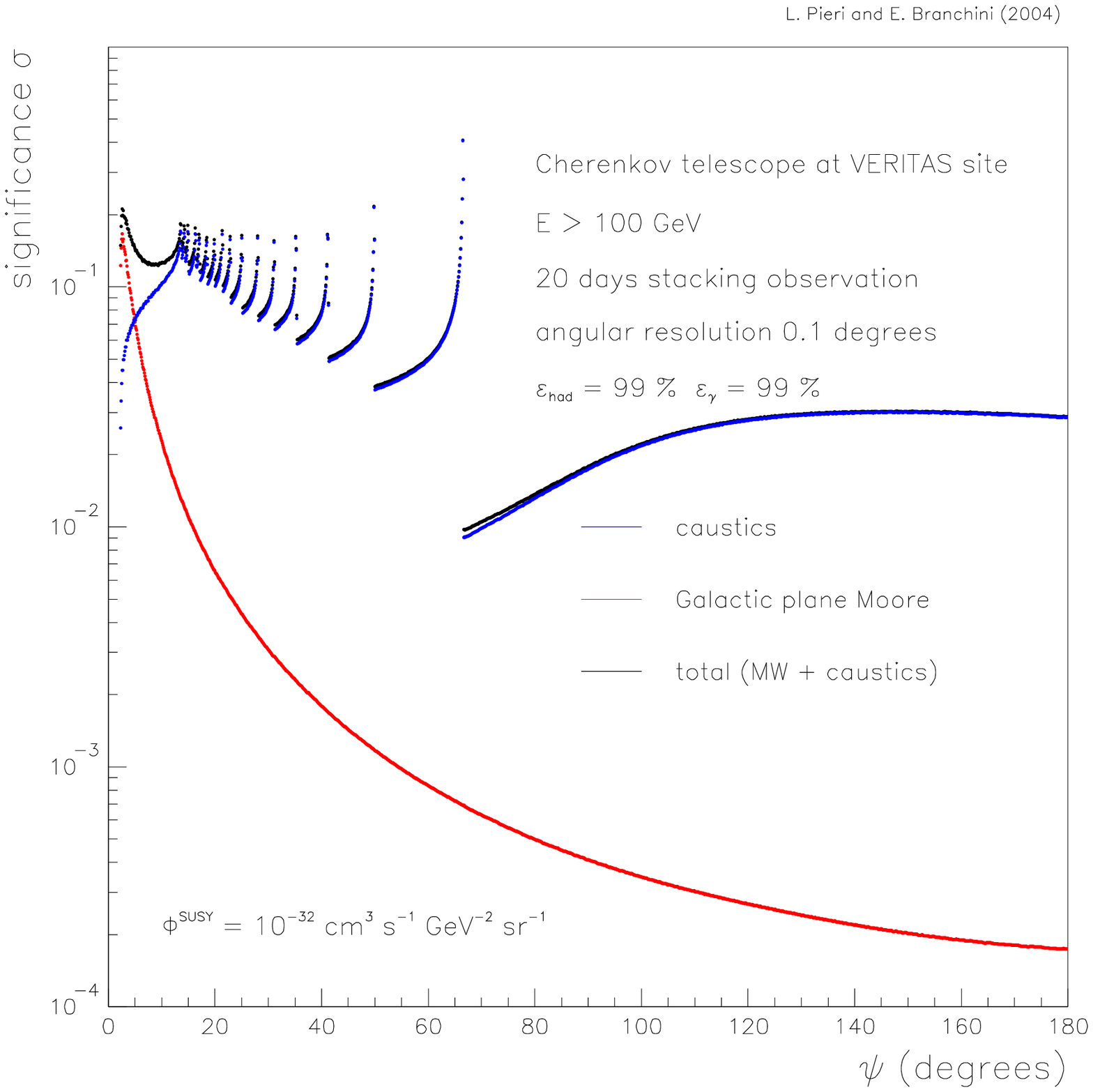}
\includegraphics[height=7.8cm,width=7.7cm]{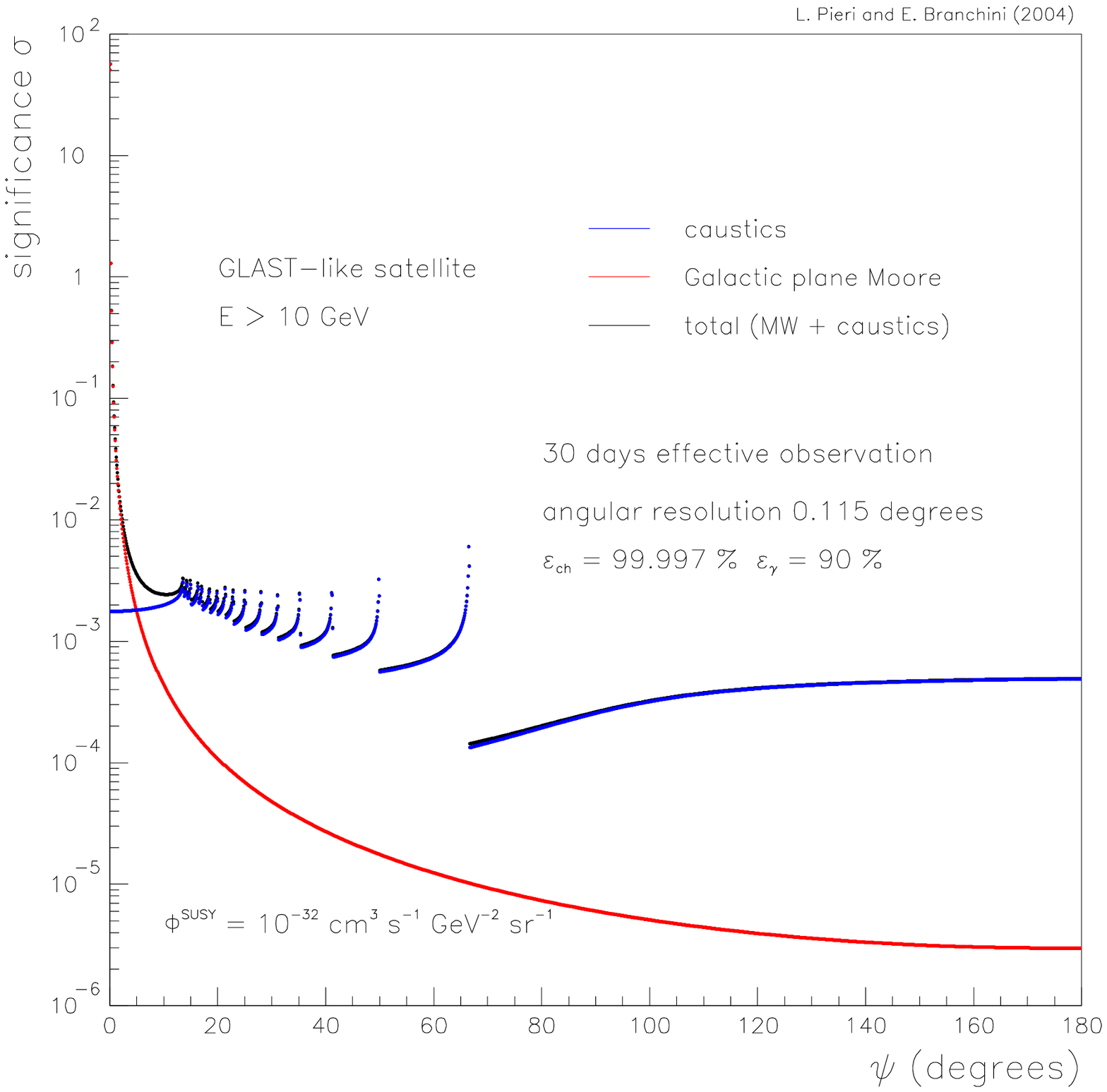}
\end{center}
\caption{Left panel: expected sensitivity of a ${\rm \check C}$erenkov 
telescope to $\gamma$-ray emission from neutralino
annihilation in the Galactic plane due to the presence of caustic rings,
at an energy threshold of 100 GeV for a solid angle of $10^{-5}$ sr. 
Right panel: the same as in the left
panel calculated for a satellite-borne experiment and an energy threshold
of 10 GeV.
\label{fig:senscau}}
\end{figure}

Fig.~\ref{fig:senscau} shows the
experimental sensitivity, $\sigma$, a function of
the angle from the GC  n the case of a ${\rm \check C}$erenkov telescope
at the VERITAS site (left panel) and  a GLAST-like satellite (right panel).
The choice of the ${\rm \check C}$erenkov telescope site 
determines the observable region of the sky and thus may constitutes a source 
For instance the GC region, which is maybe the brightest spot in the 
$\gamma$-ray sky, cannot be observed at the VERITAS site considered 
in this work.
It is worth stressing that, because of its narrow field of view,
each point in the left panels of
Fig.~\ref{fig:senscau} 
represents a single, 20-day long observation with a
${\rm \check C}$erenkov telescope.
In both panels  the significance of the detection is
is proportional to the $\gamma$-ray flux and thus
exhibits the same dependence from $\psi$
found in Fig.~\ref{fig:caustics}.\\

Clearly, the significance detection level is very low, 
even in correspondence of the peaks of the experimental sensitivity.
This means that $\gamma$-ray photons 
produced by DM annihilation in the caustics cannot be detected 
by currently planned experiments unless  
new theoretical input from particle physics could boost up
the supersymmetric factor by a factor of 30, which seems 
rather implausible.\\

These pessimistic conclusions are not alleviated when taking into 
account for the large theoretical uncertainties in the caustics model.
Indeed, as we have shown in Section 3.3, increasing the transverse 
dimension of the nearest caustics while keeping constant its distance from
the GC and using the same value of $m_\chi$,
has the effect of reducing the 
expected annihilation signal by a factor of $\sim 100$ making, it more
challenging its experimental detection.
Moreover, a further decrease of the annihilation signal is to be expected
if dynamical disturbances within the Galactic halo are large enough 
to significantly shorten their survival time (\cite{helmi:03}).

\section{Conclusions}

In the popular CDM cosmological model, high density sub-structures are 
expected to form via gravitational instability from small perturbations in the 
primordial density field of cold, collisionless particles.
In this work we have investigated the annihilation of DM particles within
caustics, that represent regions of very high density, and have estimated 
the possibility of detecting the $\gamma$-ray photons produced 
by annihilation events within our Galaxy.
Our main assumptions concern the geometry of the inner caustics, that 
we have considered small enough to be described as a series of line rings 
lying in the Galactic plane at various distances from the GC, 
as well as the underlying particle physics, namely the value
of the neutralino mass $m_\chi = 1 \TeV$ and of  
$\Phi^{\rm SUSY} = 10^{-32}  \cm^3 \sec^{-1} \GeV^{-2} \sr^{-1}$. \\

Our main conclusion is that the expected annihilation 
signal from inner Galaxy caustics cannot be revealed with 
presently available or currently planned ${\rm \check C}$erenkov
telescopes and satellite-borne detectors. 
This result takes into account the present uncertainties in
modeling the caustics geometry and their spatial distribution,
both of which systematically contribute in suppressing the expected 
annihilation signal.
As varying $m_{\chi}$ 
does not change significantly the value of
$\Phi^{\rm cosmo}$, the only possibility of observing an annihilation
signal from DM caustics would be that of assuming a much larger value
for $\Phi^{\rm SUSY}$. For example, \cite{BEG,GT} found that a 
significant detection is indeed possible in one year of observation with GLAST
if a very optimistic value of
$\Phi^{\rm SUSY}=10^{-29} \cm^3 \sec^{-1} \GeV^{-2} \sr^{-1}$ is assumed 
that, however, we regard as rather implausible.

Although the possibility of its detection is small, some additional
considerations about the annihilation signal are worth being made.
First of all, the annihilation flux from Galaxy caustics dominates
over the smooth Galactic background for 
$\psi>15^{\circ}$ and possibly also within this region
if a constant density core, rather then a sharp cusp, is present
in the GC.
Moreover, as we have checked, the annihilation signal from DM caustics
does not significantly contribute to the total annihilation signal 
expected from extragalactic sources such as M31.

\begin{figure}
\begin{center}
\includegraphics[height=7.8cm,width=7.7cm]{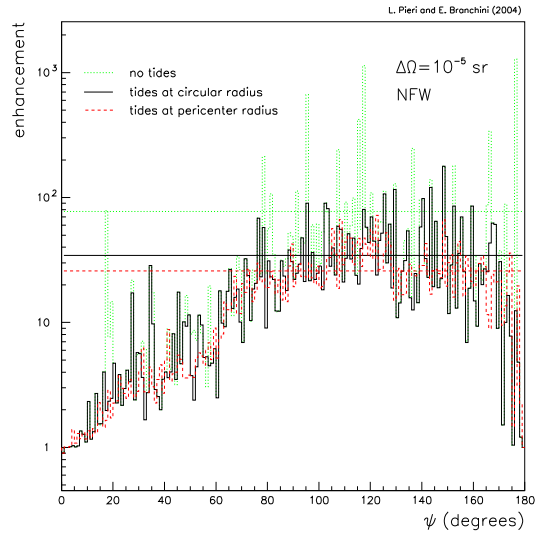}
\end{center}
\caption{Enhancement of the $\gamma$-ray emission from neutralino
annihilation in the Galactic plane due to the presence of a population
of sub-halos for a
solid angle of $10^{-5}$ sr  and a NFW97  profile.
The corresponding average values of the enhancement factor are 
indicated by the three straight lines.
\label{fig:enhsub}}
\end{figure}

Finally, it is interesting to compare the annihilation flux
in the caustics scenario with that expected from a population
of virialized sub-galactic halos that has already 
been studied by \cite{Hayashi:02,helmi:03,gao:04,DM:05,ZT:05}.
Indeed, the annihilation signal produced in Galaxy caustics
is expected to have an angular distribution across the sky 
that is very different from that produced in the smooth Galactic 
halo (Fig.~\ref{fig:caustics}) and from that 
emitted by a population of virialized sub-galactic halos.
In Fig.~\ref{fig:enhsub} we show the enhancement factor
within a solid angle of $10^{-5}$ sr
due to the presence of sub-Galactic halos 
of masses $\ge 10^{6} \msun$ containing 20\% of the original
halo mass and having the same NFW97 density  profile as the Galactic halo.
The enhancement factor obviously depends on the 
modeling of the sub-Galactic halos density profile, mass function
and spatial distribution (e.g. the three histograms
of Fig.~\ref{fig:enhsub} show the effect of using the different models of
sub-halos tidal disruptions proposed by \cite{PB:04}). Moreover,
we would like to point out again that the use of a cored subhalos' density 
profile instead of the proposed NFW97 would further decrease the expected 
fluxes. 
Yet, as we have verified, accounting for model uncertainties does not 
significantly affect the angular distribution of the annihilation signal, 
in the sense that the enhancement profile has a peak that is broader 
than that predicted by the caustics scenario with a 
maximum  at  larger angles from the GC ($\psi \sim 120^\circ$
instead of $\sim 60^\circ$).
The enhancement level remains low too, which means that 
$\gamma$-ray emission from sub-Galactic halos will hardly be detected
by future gamma-ray detectors, especially because 
observations by satellite-borne experiments like GLAST 
will be hampered by the very long 
exposure time required by the serendipitous nature 
of these $\gamma$-ray sources.

Based on these considerations, the 
GC remains the best place to look at for 
detecting $\gamma$-ray signal, 
as it is shown in Fig. \ref{fig:senscau}, provided that a sharp DM cusp
is present there.
However, ground-based observatories located in the northern hemisphere
could only observe the GC at a large zenith angle, hence introducing 
large observational errors.
For ${\rm \check C}$erenkov telescopes located in the
southern hemisphere, such as HESS and CANGAROO, as well as for
satellite-borne detectors,
the GC represents by far the brightest source of the $\gamma$-ray sky and
will therefore give the best chance to detect a possible 
signature of neutralino
annihilation. Much more sensitive detectors will  be needed to 
investigate the nature of the DM sources responsible for such signal.

\section*{Acknowledgments}
We warmly thank L. Bergstr\"om, C. Gunnarsson,
 S. Matarrese and P. Sikivie for very  useful discussions
and suggestions.
LP acknowledges a Research Grant funded 
jointly by the Italian Ministero dell'Istruzione, dell'Universit\`a e 
della Ricerca (MIUR) and by the University of Torino, and thanks the 
Department of Theoretical Physics of the University of Torino where this
work has been partially carried on. \\

\end{document}